\documentstyle[12pt]{article}

\begin{document}

\author{A. de Souza Dutra\thanks{%
E-mail: dutra@feg.unesp.br}, M. B. Hott \\
UNESP-Campus de Guaratinguet\'a-DFQ\\
Av. Dr. Ariberto Pereira Cunha, 333\\
P. O. Box 205\\
12516-410 Guaratinguet\'{a}, SP, Brasil \\
\medskip \\
V. G. C. S. dos Santos\\
Departamento de F\'{\i}sica, Centro Tecnol\'{o}gico da Aeron\'{a}utica,\\
Centro T\'{e}cnico Aeroespacial,12.228-900, \\
S. J. Campos, SP, Brasil}
\title{{\LARGE Non-Hermitian time-dependent quantum systems with real energies}}
\maketitle

\begin{abstract}
In this work we intend to study a class of time-dependent quantum systems
with non-Hermitian Hamiltonians, particularly those whose Hermitian
counterpart are important for the comprehension of posed problems in quantum
optics and quantum chemistry, which consists of an oscillator with
time-dependent mass and frequency under the action of a time-dependent
imaginary potential. The propagator for a general time-dependence of the
parameters and the wave-functions are obtained explicitly for constant
frequency and mass and a linear time-dependence in the potential. The
wave-functions are used to obtain the expectation value of the Hamiltonian.
Although it is neither Hermitean nor PT symmetric, the case under study
exhibits real values of energy.
\end{abstract}

\newpage

In the last few years, the so called PT symmetric systems introduced in the
seminal paper of Bender and Boettcher \cite{bender1} has attracted much
attention. They consist of non-Hermitian Hamiltonians with real eigenvalues,
which however exhibit parity and time-reversal symmetries or , in other
words, when one makes the space-time inversion $(P:x\rightarrow
-\,x;\,T:t\rightarrow -t,\,i\rightarrow -\,i)$. In fact, following reference
\cite{merzbacher}, the time-reversal operation $T\,\psi \left( x,t\right)
=\psi ^{*}\left( x,-t\right) $, keeps the Schroedinger equation covariant
even for time-dependent Hamiltonians \cite{wigner}, \cite{chern}. Besides,
there exist other classes of Hamiltonians having real spectra without being
PT symmetric, as can be seen, for instance, in references \cite{mostafazadeh}%
-\cite{ahmed}, and systems where the PT symmetry is spontaneously broken
\cite{bender2}, leading to complex energy eigenvalues.

Most of the papers dedicated to the subject refers to time-independent
systems and stationary states. Despite an enormous list of works devoted to
the developing and the understanding of this new kind of situation [1,
5-12], as far as we know, no one has discussed the case of time-dependent
systems, which are very important for the comprehension of many problems in
quantum optics, quantum chemistry and others areas of physics \cite{leach1}-%
\cite{castro03}. In particular we can cite the case of the electromagnetic
field intensities in a Fabry-P\'{e}rot cavity \cite{abdalla}. In fact this
kind of problem constitutes a line of investigation which still attracts the
interest of physicists \cite{guedes}-\cite{castro03}.

Here we intend to solve a class of time-dependent quantum system with
non-Hermitian Hamiltonians, particularly those which consist of an
oscillator with time-dependent mass and frequency, plus a linear
non-Hermitian term with time-dependent coupling. After that, we will discuss
the effect of that time-dependence on the reality of the spectrum, versus
the PT symmetry, by treating a particular case. The system to be studied is
represented by the one-dimensional non-Hermitian Hamiltonian, namely
\begin{equation}
H=\frac{\,p^{2}}{2\,m\left( t\right) }\,+\frac{m\left( t\right) \,\omega
^{2}\left( t\right) }{2}x^{2}\,+\,i\,\lambda \left( t\right) \,x.
\end{equation}

\noindent which in principle can preserve the PT symmetry, depending on the
time-dependence of their parameters. In fact it is easy to realize that the
PT symmetry imposes that $\lambda \left( t\right) $ must be an even function
of time for $m\left( t\right) $ and $\omega ^{2}\left( t\right) $ symmetric
under time-reversal.

The corresponding Schroedinger equation is written as
\begin{equation}
-\frac{1}{2\,m\left( t\right) }\frac{\partial ^{2}\psi \left( x,t\right) }{%
\partial x^{2}}\,+\,\left[ \frac{1}{2}\,m\left( t\right) \,\omega \left(
t\right) ^{2}\,x^{2}\,+\,i\,\lambda \left( t\right) \,x\right] \psi \left(
x,t\right) \,=\,i\,\frac{\partial \psi \left( x,t\right) }{\partial t},
\label{scheq}
\end{equation}

\noindent where $m\left( t\right) $, $\omega \left( t\right) $ and $\lambda
\left( t\right) $ are time-dependent functions (We set $\hbar \,=\,1$ for a
while).

Usually the time-independent forced harmonic oscillator is solved by means
of a translation of the coordinates. When we deal with a time-dependent
forced harmonic oscillator one can check that a trivial translation of the
coordinates $x=y-i\frac{\,\lambda \left( t\right) }{2m\left( t\right)
\,\omega ^{2}\left( t\right) }$ does not lead to a harmonic oscillator with
a time-dependent mass and frequency because we are still left with a first
order derivative term in $y$ and an additional purely time-dependent term
(see Eq. (5) below).

To show that the Schroedinger equation (\ref{scheq}) can be mapped into an
evident exactly solvable equation we appeal to a generalization of
transformations, which envolves a change of variable and time
reparametrization, used before in several time-dependent systems \cite{Cheng}%
, \cite{fara}, \cite{dutra}, namely

\begin{equation}
x\,=\,s\left( \tau \right) \,y\,+\,i\,\eta \left( \tau \right) ,
\label{transfofvar1}
\end{equation}

\noindent where $\tau $ is a single-valued function related to the original
time $t$ by
\begin{equation}
\tau \left( t\right) \,=\,\int^{t}\mu \left( \xi \right) \,d\xi
,\,\,\,\,\,\,\,\,\frac{d\tau \left( t\right) }{dt}\,=\,\mu \left( t\right) .
\label{transofvar2}
\end{equation}

\noindent In the transformation (\ref{transfofvar1}), $\eta $ is a real
function and $y$ is necessarily complex to keep the original physical
variable $x$ real. After the transformation of variables we are led to the
following equation
\[
i\,\mu \,\frac{\partial \psi \left( y,\tau \right) }{\partial \tau }\,-\,i%
\frac{\,\mu }{s}\,\left[ \acute{s}\,\,y\,+\,i\,\acute{\eta}\,\right] \frac{%
\partial \psi \left( y,\tau \right) }{\partial y}\,+\,\frac{1}{2\,m\,s^{2}}%
\frac{\partial ^{2}\psi \left( y,\tau \right) }{\partial y^{2}}-
\]
\begin{equation}
-\,\left[ \frac{1}{2}m\,\omega ^{2}(s^{2}y^{2}+2\,i\,s\,\eta \,y-\eta
^{2})+i\,\lambda \,s\,y-\lambda \,\eta \right] \psi \left( y,\tau \right) =0.
\end{equation}

\noindent At this point, in order to get the Schroedinger equation in the
new variables, we introduce a wavefunction redefinition
\begin{equation}
\psi \left( y,\tau \right) =\,\exp \left[ i\,f\left( y,\tau \right) \right]
\sigma \left( y,\tau \right) ,  \label{wave}
\end{equation}

\noindent which once substituted in the above equation gives us
\[
\left\{ i\,\mu \,\frac{\partial }{\partial \tau }+\frac{1}{2\,m\,s^{2}}\frac{%
\partial ^{2}}{\partial y^{2}}-\frac{1}{2}m\,\omega ^{2}s^{2}\left[ y^{2}+%
\frac{2\,i}{s}\left( \eta +\frac{\lambda }{m\omega ^{2}}\right) y-\frac{\eta
}{s^{2}}\left( \eta +\frac{2\lambda }{m\omega ^{2}}\right) \right] +\right.
\]
\[
\left. +\frac{1}{2ms^{2}}\left[ i\frac{\partial ^{2}f}{\partial y^{2}}%
-\left( \frac{\partial f}{\partial y}\right) ^{2}\right] +\mu \,\frac{\acute{%
s}}{s}\,y\frac{\partial f}{\partial y}+i\frac{\acute{\eta}\,\mu }{m\omega
^{2}s}\frac{\partial f}{\partial y}-\mu \frac{\partial f}{\partial \tau }%
\right\} \sigma \left( y,\tau \right) +
\]

\begin{equation}
+\left[ \;i\frac{1}{m\,s^{2}}\frac{\partial f}{\partial y}-i\,\mu \,\frac{%
\acute{s}}{s}\,y+\frac{\acute{\eta}\,\mu }{s}\right] \frac{\partial \sigma
\left( y,\tau \right) }{\partial y}=0,
\end{equation}

\noindent where the prime denotes differentiation with respect to $\tau $.
Now we choose conveniently the arbitrary function $\,f(y,\tau )$ to
guarantee that the coefficient of the term $\frac{\partial \sigma }{\partial
y}$ vanishes identically. In doing so we get
\begin{equation}
\frac{\partial f}{\partial y}=m\,\mu \,s\,\acute{s}\,\,y+ims\mu \,\acute{\eta%
},
\end{equation}

\noindent which has as a general solution
\begin{equation}
f\,(y,\tau )=\,\frac{1}{2}\,m\,\mu \,s\,\acute{s}\,\,y^{2}\,+ims\mu \,\acute{%
\eta}y+\,f_{\tau }\left( \tau \right) ,
\end{equation}

\noindent with $f_{\tau }\left( \tau \right) $ an arbitrary function of the
rescaled time $\tau $, which is still to be determined by some convenient
imposition. Now, substituting this function into the equation (7), one can
rewrite it as
\begin{eqnarray*}
&&\left\{ i\,\mu \,\frac{\partial }{\partial \tau }+\frac{1}{2\,m\,s^{2}}%
\frac{\partial ^{2}}{\partial y^{2}}-\frac{\mu }{2}\left[ \frac{m\,\omega
^{2}s^{2}}{\mu }+\frac{d}{d\tau }\left( m\,\mu \,s\,\acute{s}\right) -\mu m\,%
\acute{s}^{2}\,\right] y^{2}+\right. \\
&&\left. -i\,\mu \left[ \frac{m\omega ^{2}s}{\mu }\left( \eta +\frac{\lambda
}{m\omega ^{2}}\right) +\frac{d}{d\tau }\left( ms\mu \,\acute{\eta}\right)
\right] y+\right.
\end{eqnarray*}
\[
\left. +\frac{i}{2}\mu \frac{\acute{s}}{s}\,-\mu \frac{\partial f_{\tau }}{%
\partial \tau }+\frac{1}{2}m\omega ^{2}\eta \left( \eta +\frac{2\lambda }{%
m\omega ^{2}}\right) -\frac{m\mu ^{2}\,\acute{\eta}^{2}}{2}\right\} \sigma
\left( y,\tau \right) =0.
\]

\noindent We choose the $f_{\tau }\left( \tau \right) $ and impose some
constraints on the functions $s(\tau )$, $\eta (\tau )$ and $\mu (\tau )$ in
order to leave the effective potential depending only on $y^{2}$. That is $%
f_{\tau }\left( \tau \right) $ must to be such that
\begin{equation}
\frac{\partial f_{\tau }}{\partial \tau }=\frac{1}{2}\left[ \frac{i\acute{s}%
}{s}\,+\frac{m\omega ^{2}\eta }{\mu }\left( \eta +\frac{2\lambda }{m\omega
^{2}}\right) -\mu \,m\,\acute{\eta}^{2}\right] ,  \label{constraint1}
\end{equation}

\noindent whose solution can be cast into the form below without loss of
generality
\begin{equation}
f_{\tau }\left( \tau \right) =\,i\,\ln \left( s^{\frac{1}{2}}\right) +\,%
\frac{1}{2}\int^{\tau }\left[ \frac{m(\xi )\omega (\xi )^{2}\eta (\xi )}{\mu
(\xi )}\left( \eta (\xi )+\frac{2\lambda (\xi )}{m(\xi )\omega (\xi )^{2}}%
\right) -\,m(\xi )\mu (\xi )\,\acute{\eta}(\xi )^{2}\right] d\xi .
\end{equation}

\noindent Besides $s(\tau )$, $\eta (\tau )$ and $\mu (\tau )$ have to
satisfy the equation

\begin{equation}
\frac{d}{d\tau }\left( ms\mu \,\acute{\eta}\right) =-\frac{m\omega ^{2}s}{%
\mu }\left( \eta +\frac{\lambda }{m\omega ^{2}}\right) .  \label{constraint2}
\end{equation}

\noindent which allow us to write the corresponding Schroedinger equation in
the form
\begin{equation}
\left[ i\,\mu \,\frac{\partial }{\partial \tau }+\frac{1}{2\,m\,s^{2}}\frac{%
\partial ^{2}}{\partial y^{2}}-\frac{1}{2}m\,s^{2}\left( \omega ^{2}+\Omega
^{2}\right) y^{2}\right] \sigma \left( y,\tau \right) =0,
\end{equation}

\noindent where
\begin{equation}
\Omega ^{2}\equiv \,\frac{\mu }{m\,s^{2}}\frac{d}{d\tau }\left( m\,\mu \,s\,%
\acute{s}\right) -\,\left( \mu \frac{\acute{s}}{s}\right) ^{2}.
\end{equation}
Until now, the functions $s\left( \tau \right) $ and $\mu \left( \tau
\right) $ remains arbitrary since equation (\ref{constraint2}) constrains $%
\eta (\tau )$ in terms of $s\left( \tau \right) $ and $\mu \left( \tau
\right) $. Then we use this arbitrariness to get a simpler equation to this
problem. One obvious possibility is choosing them in order to get a
time-independent Schroedinger equation for $\sigma \left( y,\tau \right) $%
\begin{equation}
\mu \left( i\,\frac{\partial }{\partial \tau }+\frac{1}{2\,m_{0}}\frac{%
\partial ^{2}}{\partial y^{2}}-\frac{1}{2}m_{0}\,\omega
_{0}^{2}\,\,y^{2}\right) \sigma \left( y,\tau \right) =0,  \label{eqforsigma}
\end{equation}

\noindent where it can be seen that the time dependence was factorized,
implying that
\begin{equation}
m\,s^{2}\mu =\,m_{0}=\,const.;\,\,\frac{m\,s^{2}}{\mu }\left( \omega
^{2}+\Omega ^{2}\right) =m_{0}\,\omega _{0}^{2}\,=const.\,\,
\label{nonlinear}
\end{equation}

From the above equations one can see that we have transformed the original
problem into that of an usual harmonic oscillator with constant mass and
frequency. At this point we are able to calculate the propagator for the
system, once it is a special solution of the Schroedinger equation, subject
to the restriction
\begin{equation}
\lim_{t\rightarrow t_{0}}K\left( x,t;x_{0},t_{0}\right) \,=\,\delta \left(
x-x_{0}\right) .
\end{equation}

\noindent Now, remembering that in this case the wavefunction is written as
\begin{equation}
\psi \left( x,t\right) \,=\,\left. e^{i\,f\left( y,\tau \right) }\,\sigma
\left( y,\tau \right) \right| _{y=x/s\left( \tau \right) -i\,\eta \left(
\tau \right) /s\left( \tau \right) ,\,\tau =\tau \left( t\right) },
\end{equation}

\noindent and finally using that
\begin{equation}
\psi \left( x,t\right) =\,\int_{-\infty }^{\infty }K\left(
x,t;x_{0},t_{0}\right) \,\psi \left( x_{0},t_{0}\right) ,
\end{equation}

\noindent it is easy to conclude that the propagator will come from \cite
{fara}
\begin{equation}
K\left( x,t;x_{0},t_{0}\right) =\,\left. e^{i\,f\left( y,\tau \right)
}\,K\left( \bar{x},\tau ;\bar{x}_{0},\tau _{0}\right) \,\,e^{-i\,f^{*}\left(
y_{0},\tau _{0}\right) }\right| _{y=\left( x\,-i\,\eta \left( \tau \right)
\right) /s\left( \tau \right) ,\,\tau =\tau \left( t\right) }\,\,,
\end{equation}

\noindent where $\,f^{*}\left( y_{0},\tau _{0}\right) $ stands for the
complex conjugate evaluated at initial time and coordinate. Now using the
well known expression of the propagator of the harmonic oscillator \cite
{feynman}, one obtains
\[
K\left( x,t;x_{0},t_{0}\right) =\left\{ \frac{m_{0}\,\omega _{0}}{2\,\pi
\,i\,\hbar \,s\,s_{0}\,\sin \left[ \omega _{0}\left( \tau -\tau _{0}\right)
\right] }\right\} ^{1/2}\exp \frac{i}{2\,\hbar }\left\{ \frac{m\,\dot{s}%
\,y^{2}}{s}-\frac{m\,\dot{s}_{0}}{s_{0}}\,y_{0}^{2}+\right. \,
\]
\begin{equation}
\left. +\,\,\frac{\,m_{0}\,\omega _{0}}{\sin \left[ \omega _{0}\left( \tau
-\tau _{0}\right) \right] }(y^{2}++y_{0}^{2})\cos \left[ \omega _{0}\left(
\tau -\tau _{0}\right) \right] -2\,y\,\,y_{0}\right\} _{\,y=\left(
x\,-i\,\eta \left( t\right) \right) /s\left( \tau \right) ,\,\tau =\tau
\left( t\right) }\,,
\end{equation}

\noindent where it can be observed that we recall the dependence on $\hbar $%
, and use the identification $\mu \,s^{\prime }=\dot{s}$, where the dot
denotes differentiation with respect to $t$. Furthermore one has a nonformal
solution provided that an exact solution of (\ref{constraint2}) and (\ref
{nonlinear}) is known \cite{Cheng}-\cite{guedes},\cite{castro03}.

Let us now obtain an expression for the wavefunctions directly from the
above propagator. For this we make use of the Mehler's formula \cite{mehler}
\begin{equation}
\frac{\exp \left[ -\frac{\left( a^{2}+b^{2}-2\,a\,b\,c\right) }{\sqrt{1-c^{2}%
}}\right] }{\sqrt{1-c^{2}}}=\exp \left[ -\left( a^{2}+b^{2}\right) \right]
\sum_{n=0}^{\infty }\frac{c^{n}}{2^{n}n!}\,H_{n}\left( a\right)
\,H_{n}\left( b\right) ,
\end{equation}

\noindent with $a\equiv \sqrt{\frac{m_{0}\,\omega _{0}}{\hbar }}\,y_{0}$, $b=%
\sqrt{\frac{m_{0}\,\omega _{0}}{\hbar }}\,y$, $c=\exp \left\{ -i\left[
\omega _{0}\left( \tau -\tau _{0}\right) \right] \right\} $. Moreover, the
propagator can be written in its spectral decomposition form
\begin{equation}
K\left( y,\tau ;y_{0},\tau _{0}\right) =\sum_{n=0}^{\infty }\psi
_{n}^{*}\left( y,\tau \right) \psi \left( y_{0},\tau _{0}\right) ,
\end{equation}

\noindent and from which the wavefunctions can be devised as
\[
\psi _{n}\left( y,\tau \right) =\exp \left[ -i\left( n+\frac{1}{2}\right)
\mu \left( t\right) \right] \,\phi _{n}\left( y,\tau \right) ,
\]
\begin{eqnarray}
\phi _{n}\left( y,\tau \right)  &=&\sqrt{\frac{1}{2^{n}n!}\sqrt{\frac{%
m_{0}\,\omega _{0}}{\pi \,\hbar \,s}}}\,\exp \left\{ -\,\frac{\,1}{2\,\hbar }%
\left[ m_{0}\,\omega _{0}+i\,\frac{m\left( \tau \right) \dot{s}}{s}\right]
y^{2}\right\} \,\,.  \nonumber \\
&&\left. H_{n}\left( \sqrt{\frac{m_{0}\,\omega _{0}}{\,\hbar }}\,\,y\right)
\right| _{y=\left( x\,-i\,\eta \left( t\right) \right) /s\left( \tau \right)
,\,\tau =\tau \left( t\right) }\,\,.
\end{eqnarray}

\noindent In the above expressions $H_{n}$ stands for the Hermite polynomial.

\smallskip Let us now treat a simpler particular case, which can be used to
get a deep understanding of the problem without unnecessary mathematical
complications namely, the case of a usual harmonic oscillator with a
time-dependent PT-violating linear potential,
\begin{equation}
-\frac{\hbar ^{2}}{2\,m}\frac{\partial ^{2}\psi \left( x,t\right) }{\partial
x^{2}}\,+\,\left[ \frac{1}{2}\,m\,\,\omega ^{2}\,x^{2}\,+\,i\,at\,x\right]
\psi \left( x,t\right) \,=\,i\,\hbar \,\frac{\partial \psi \left( x,t\right)
}{\partial t},
\end{equation}

\noindent where $a$ is an arbitrary real constant. The coordinate and time
transformations given in equations (\ref{transfofvar1}) and (\ref
{transofvar2}) can be simplified by choosing $s(\tau )=\mu (\tau )=1,$ $%
\left( \tau =t\right) $ and $\eta (t)=-\frac{a\,t}{m\omega ^{2}}$. Then one
ends up with the Schroedinger equation (\ref{eqforsigma}) where $\omega
_{0}=\omega $ and $m_{0}=m$, whose solutions in the transformed coordinate
looks like simply as
\begin{equation}
\,\sigma _{n}\left( y,t\right) =\,\sqrt{\frac{1}{2^{n}n!}\sqrt{\frac{%
m\,\,\omega }{\pi \,\hbar \,}}}\,e^{-\frac{m\,\omega }{2\,\hbar }%
\,y^{2}-\,i\,\left( n+\frac{1}{2}\right) \omega \,t}\,H_{n}\left( \sqrt{%
\frac{m\,\omega }{\,\hbar }}\,\,y\right) .
\end{equation}

At this point we are able to calculate the energy expectation value of the
system. For this we calculate, as usual, the expression
\begin{equation}
\langle E\rangle \,=\frac{\,\int_{-\,\infty }^{\infty }dx\,\psi ^{*}\left(
x,t\right) \,i\,\hbar \,\frac{\partial \psi \left( x,t\right) }{\partial t}}{%
\,\int_{-\,\infty }^{\infty }dx\,\psi ^{*}\left( x,t\right) \,\psi \left(
x,t\right) \,}\,.
\end{equation}

\noindent After straigthforward calculations one can get

\begin{equation}
\ \langle E\rangle \,=\left( n+\frac{1}{2}\right) \hbar \,\omega \,-\hbar \,%
\dot{\gamma}(t)-m\omega \,\dot{\eta}\left[ \frac{\,\int_{-\,\infty }^{\infty
}dx\,\psi ^{*}\left( x,t\right) \left( \,y+2\frac{\,H_{n-1}(y)}{\,H_{n}(y)}%
\right) \psi \left( x,t\right) }{\,\int_{-\,\infty }^{\infty }dx\,\psi
^{*}\left( x,t\right) \,\psi \left( x,t\right) \,}\right] _{y=x\,-i\,\eta
\left( t\right) },
\end{equation}

\noindent where
\[
\gamma (t)=\frac{a^{2}t}{2\hbar m\omega ^{4}}\left( 1-\frac{\omega ^{2}t^{2}%
}{3}\right) .
\]

We do not have a general expression for the energy expectation value for an
arbitrary eigenstate, but we have verified that they are always real. In
fact, the calculation for a particular state can be done straightforwardly
and we have, for instance

\begin{eqnarray*}
\langle 0\left| E\right| 0\rangle &=&\frac{1}{2}\hbar \,\omega \,-\hbar \,%
\dot{\gamma}(t)-\frac{ma}{\omega }\,\dot{\eta},\,\,\,{\rm and} \\
\langle 1\left| E\right| 1\rangle &=&\frac{3}{2}\hbar \,\omega \,-\hbar \,%
\dot{\gamma}(t)-m\omega \,\dot{\eta}\left[ \frac{(3\hbar ^{2}\omega
^{4}+2a^{2})-2a\hbar ^{3/2}\omega ^{3/2}m^{-1/2}}{(\hbar ^{2}\omega
^{4}+2a^{2})}\right] .
\end{eqnarray*}

This way we have proven that the reality of the time-dependent energy is
mantained despite of the non-Hermiticity of the Hamiltonian. For a
non-Hermitian and PT-violating Hamiltonian this is a very intriguing
conclusion and demands further investigations on the origin of the reality
of the energy in such systems, which can also be strongly considered as
candidates for realizations of non-Hermitian interactions. Furthermore, one
can note that even with the mapping into a Hermitian time-independent
system, Eq. (\ref{eqforsigma}), it is not trivial to verify that the
original time-dependent system has real energies. In fact, it is necessary
to use the wavefunction (\ref{wave}) to calculate the energy expectation
value. We are still studying cases where the mass and frequency are
constants, looking for the more general time-dependent forces, including PT\
symmetric ones, that present real spectrum.

\bigskip

\noindent {\bf Acknowledgments:} This work was partially supported by
FAPESP, CNPq and CAPES.

\end{document}